\documentclass[aps,prl,floatfix,twocolumn,tightenlines,superscriptaddress]{revtex4}

\usepackage[bookmarks = true, colorlinks = true, urlcolor=blue]{hyperref}
\usepackage[utf8x]{inputenc}
\usepackage{amsmath}
\usepackage{latexsym}
\usepackage{graphicx}
\usepackage{color}
\usepackage{bm}

\newcommand{\ket}[1]{|#1\rangle}

\newcommand{\SH}{\mathbf{H}}
\newcommand{\SV}{\mathbf{V}}
\newcommand{\SD}{\mathbf{D}}
\newcommand{\SA}{\mathbf{A}}
\newcommand{\SL}{\mathbf{L}}
\newcommand{\SR}{\mathbf{R}}

\begin{document}

\title{Minimum-error discrimination of entangled quantum states}
\author{Y. Lu}
\affiliation{Institute for Quantum Computing and Department of
Physics \& Astronomy, University of Waterloo, Waterloo, Canada,
N2L 3G1}
\author{N. Coish}
\affiliation{Institute for Quantum Computing and Department of
Physics \& Astronomy, University of Waterloo, Waterloo, Canada,
N2L 3G1}
\author{R. Kaltenbaek}
\affiliation{Institute for Quantum Computing and Department of
Physics \& Astronomy, University of Waterloo, Waterloo, Canada,
N2L 3G1}
\author{D.R. Hamel}
\affiliation{Institute for Quantum Computing and Department of
Physics \& Astronomy, University of Waterloo, Waterloo, Canada,
N2L 3G1}
\author{S. Croke}
\affiliation{Institute for Quantum Computing and Department of
Physics \& Astronomy, University of Waterloo, Waterloo, Canada,
N2L 3G1}
\affiliation{Perimeter Institute for Theoretical Physics, Waterloo,
Canada, N2L 2Y5}
\author{K.J. Resch}
\email{kresch@iqc.ca}
\affiliation{Institute for Quantum Computing and Department of
Physics \& Astronomy, University of Waterloo, Waterloo, Canada,
N2L 3G1}

\begin{abstract}
\noindent Strategies to optimally discriminate between quantum
states are critical in quantum technologies.  We present an
experimental demonstration of minimum error discrimination
between entangled states, encoded in the polarization of pairs
of photons.  Although the optimal measurement involves
projecting onto entangled states, we use a result of Walgate
\emph{et al.} to design an optical implementation employing
only local polarization measurements and feed-forward, which
performs at the Helstrom bound.  Our scheme can achieve perfect
discrimination of orthogonal states and minimum error
discrimination of non-orthogonal states. Our experimental
results show a definite advantage over schemes not using
feed-forward.

\end{abstract}

\maketitle

The ability, or inability, to perfectly distinguish between
different quantum states is a defining feature of quantum
mechanics and the basis of many quantum technologies. A typical
scenario is that one wants to distinguish between one of two
possible quantum states. If the two states are orthogonal, then
there exists a measurement that can always distinguish between
them with certainty. While there is no such test for
non-orthogonal states, there are many ways to quantify how well
a measurement distinguishes between them, and each defines a
different strategy \cite{Chefles00, Barnett09, Helstrom76,
Holevo73, Yuen75, Ivanovic87, Dieks88, Peres88, Davies78,
Sasaki99, Croke06}. The most commonly considered are the
minimum error \cite{Helstrom76,Holevo73,Yuen75} and unambiguous
discrimination strategies \cite{Ivanovic87,Dieks88,Peres88};
the former arises naturally as an extension of classical
hypothesis testing to the quantum realm while the latter is a
peculiarly quantum strategy where error-free discrimination may
sometimes be achieved by allowing for the possibility of an
inconclusive outcome.

Each strategy is defined by optimizing a specific figure of
merit over all possible measurements. However, whether the
optimal measurement is achievable in practice depends on the
physical system in question and its practical limitations.
Experimental demonstrations of quantum state discrimination
protocols to date have all used either the polarization or the
spatial paths of light to encode the quantum information. Thus
far, measurements consist of linear optics and photodetectors
to measure the intensity of light in different paths; as the
proportion of light reaching any given detector is the same for
a beam of single photons and a weak classical source, almost
all used attenuated lasers. The optimal minimum error
\cite{Barnett97,Clarke01}, unambiguous
\cite{Huttner96,Clarke01b,Mohseni04}, maximum mutual
information \cite{Barnett97,Clarke01,Mizuno01}, and maximum
confidence \cite{Mosley06} strategies were implemented in this
way. A very recent work investigated a state discrimination
strategy using heralded single photons from down-conversion
\cite{higgins09}.

\begin{figure}
\begin{center}
\includegraphics[width=1 \columnwidth]{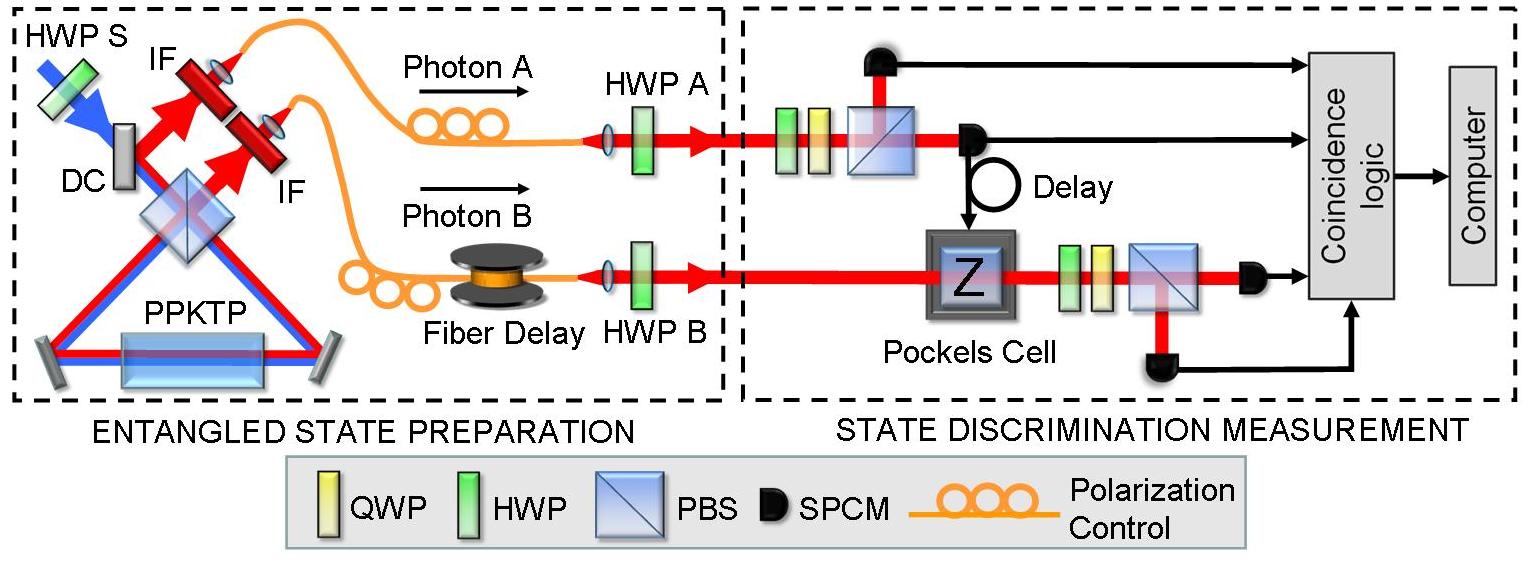}
\caption{Experimental setup.
Half-wave plates S, A, and B
were rotated to prepare polarization entangled states from the source.  Feed-forward
was achieved using a fast Pockels cell (PC) which performs the
identity when off, or a phase flip (Pauli-$Z$) when on.  We
detect photons at single-photon counting modules (SPCM) and monitor coincidence
events. With the PC off we also used this setup for quantum
state tomography.  Quarter-wave plate (QWP); polarizing
beam-splitter (PBS); periodically-poled KTP crystal (PPKTP); interference filter (IF).
See text for more details.
\label{setup}}
\end{center}
\end{figure}

Here, we experimentally demonstrate state discrimination
between pairs of \emph{entangled} states, a truly quantum
regime. We separately consider discrimination of orthogonal and
non-orthogonal states.  For the orthogonal case, one can always
perform a projective measurement onto the states themselves to
achieve perfect discrimination. In the non-orthogonal case, the
theory describing the optimal measurements for single qubits
\cite{Chefles00, Barnett09, Helstrom76, Holevo73, Yuen75,
Ivanovic87, Dieks88, Peres88, Davies78, Sasaki99, Croke06} may
readily be applied to pairs of entangled states. However,
experimental constraints may prohibit implementing these
optimal measurements. Specifically, with optical polarization
high-fidelity local polarization measurements are possible, but
entangling measurements are very difficult
\cite{Calsamiglia2001a}. One might expect these constraints to
limit our ability to discriminate between entangled states, but
this is not the case. \emph{Any} two orthogonal pure states,
irrespective of their entanglement or multipartite structure,
may be perfectly discriminated using only local operations and
classical communication (LOCC) \cite{Walgate00}. Further, any
two non-orthogonal pure states can be optimally discriminated
by LOCC, according to both the minimum error \cite{Virmani01}
and unambiguous discrimination \cite{Chen01,Chen02,Ji05}
figures of merit.  This highlights how to overcome seeming
limitations of a physical system by using its resources more
effectively. Conversely, and equally surprisingly, there exist
sets of orthogonal \emph{product} states between which perfect
discrimination is \emph{not} possible by LOCC \cite{Bennett99}.
The relationship between entanglement and LOCC discrimination
is not simple, but helps shed some light on the power and
limitations of LOCC.

The key insight for discriminating pairs of entangled states
via LOCC was developed by Walgate \emph{et al.}
\cite{Walgate00}. They showed that any two orthogonal quantum
states, $|\phi\rangle$ and $|\psi\rangle$, where Alice has
subsystem $A$ and Bob has subsystem $B$, can be written in the
form, $|\phi\rangle = \sum_i c_i|i\rangle_A|\eta_i\rangle_B$
and $|\psi\rangle = \sum_i
d_i|i\rangle_A|\eta^\perp_i\rangle_B$, where $\{|i\rangle\}$ is
a basis for subsystem $A$ and
$\langle\eta_i|\eta^\perp_i\rangle=0$.  Note that $\langle
i\vert j\rangle = \delta_{ij}$ but $\ket{\eta_i}$ and
$\ket{\eta_j}$ need not be orthogonal, differentiating this
decomposition from the more widely known Schmidt decomposition.
By viewing the states in this way, a strategy for
distinguishing them becomes clear. Alice measures in the basis
$\{|i\rangle\}$ and sends her result to Bob. Bob then orients
his measurement device to distinguish between $|\eta_i\rangle$
and $|\eta^\perp_i\rangle$. In general, Bob needs to be able to
change his measurement basis based on Alice's measurement
outcome; this is called \emph{feed-forward}.  For two
non-orthogonal pure states, the minimum-error, or Helstrom
\cite{Helstrom76}, measurement for discriminating between them
is a projective measurement, i.e., there must be two orthogonal
states that are perfectly discriminated by this measurement.
Since such a measurement can be performed by LOCC, the
minimum-error measurement to discriminate between any pair of
non-orthogonal pure states may also be performed using LOCC
\cite{Virmani01}. For pure states with more than two
subsystems, one can use LOCC for optimal discrimination by
employing this strategy iteratively \cite{Walgate00}.

To implement this protocol experimentally, we constructed the
setup shown in Fig.~\ref{setup}.  A 404.5~nm, 0.25~mW
grating-stabilized diode laser pumps a polarization-based
Sagnac interferometer, producing 809-nm polarization-entangled
photon pairs via type-II parametric down-conversion in a 25-mm
long periodically-poled KTP (PPKTP) crystal
\cite{kim06,fedrizzi07,biggerstaff09}. The down-converted
photons are coupled into single-mode fibres. When these fibres
are directly connected to our single-photon counting detectors
(Perkin Elmer SPCM-AQ4C) we typically measure coincidence rates
of 7~kHz and singles rates of 60~kHz. Using the half-wave plate
in the source (HWP S) we control the degree of polarization
entanglement of the photon pairs. When set to produce maximum
entanglement, we typically measure a fidelity
\cite{Jozsa1994a}, $F=0.97$, with the target Bell state, and
tangle \cite{Coffman00}, $\tau=0.93$. The polarization of the
photons is measured using analyzers consisting of a HWP, a
quarter-wave plate (QWP), and a polarizing beam-splitter (PBS).
Signals detected in the transmitted arm of the PBS in the
analyzer for photon 1 are delayed by 250ns and then trigger a
Pockels cell (Leysop RTP4-20-AR800) that implements a Pauli-Z
operation, $Z$, ($\ket{\SH}\rightarrow\ket{\SH}$,
$\ket{\SV}\rightarrow -\ket{\SV}$) on photon 2.

In the first part of our experiment, we investigate a set of
orthogonal two-qubit entangled states:
\begin{eqnarray}
\label{orthstates}
|\phi_0(\theta_0)\rangle&=&\cos\theta_0\ket{\SH}Z |u\rangle + \sin\theta_0 \ket{\SV} \ket{u}  \\
|\phi_1(\theta_1)\rangle&=&\cos\theta_1\ket{\SH}Z |u^\perp\rangle - \sin\theta_1 \ket{\SV} |u^\perp\rangle, \nonumber
\end{eqnarray}
\noindent where $\ket{\SH}$ ($\ket{\SV}$) represent horizontal
(vertical) polarization of a photon, $|u\rangle =
\frac{\sqrt{3}}{2}\ket{\SH}+\frac{1}{2}\ket{\SV}$,
$|u^\perp\rangle =
\frac{1}{2}\ket{\SH}-\frac{\sqrt{3}}{2}\ket{\SV}$.  While our
parametrization does not allow the generation of arbitrary
pairs of states, these pairs encapsulate the full difficulty of
discrimination as optimal performance requires feed-forward for
most values of $\theta_0$ and $\theta_1$. Note that any two
orthogonal states can always be perfectly distinguished using
local measurements \emph{without} feed-forward when the states
share no support in some product basis; this includes the cases
where at least one of the states is a product state or
maximally entangled.

\begin{figure}
\begin{center}
\includegraphics[width=0.9 \columnwidth]{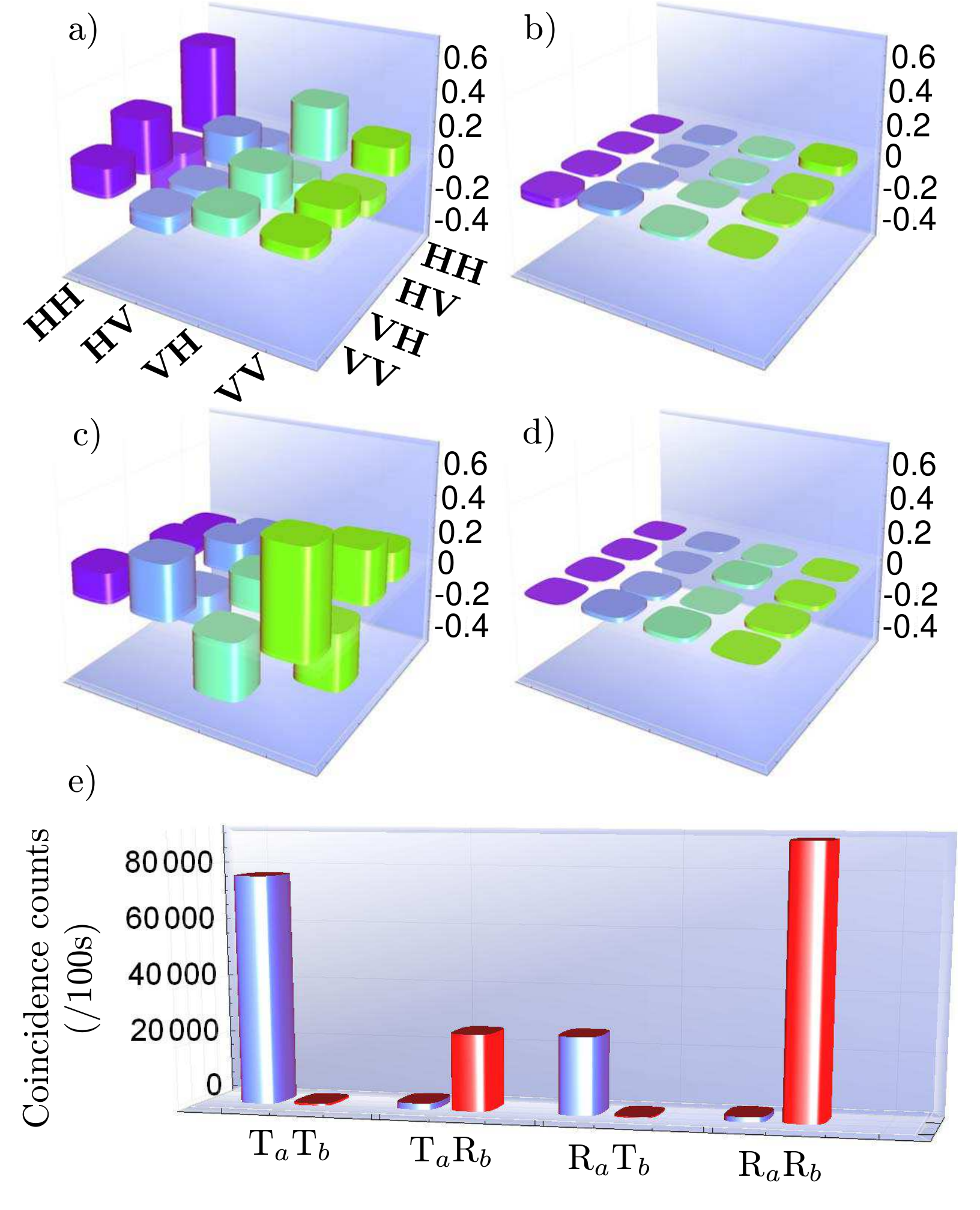}
\caption{Discrimination of two orthogonal entangled states.
We set HWPs S, A, and B to prepare the state
$\vert\phi_0\left(30^\circ\right)\rangle$ and used quantum
state tomography to measure the state produced.  The density
matrix is depicted as a bar chart in a) real part and b)
imaginary part. The fidelity with the target state is $0.981\pm0.002$.
Similarly, we prepared $\vert\phi_1\left(60^\circ\right)\rangle$ with a
fidelity of $0.986\pm0.002$ and show the reconstructed state in c) real
part and d) imaginary part. In e) we show the measured coincidence
counts for the state-discrimination measurement with feed-forward for
the state from a) \& b) (blue, left bars) and the state from c) \& d)
(red, right bars); from these counts we find that the probability for
distinguishing these states is $0.9780\pm0.0003$, significantly exceeding the success
probability for the best possible projective measurement without
feed-forward, $0.922\pm0.002$.
\label{sampletomocounts}}
\end{center}
\end{figure}

We used HWPs S, A, and B to prepare the states in
Eq.~\ref{orthstates} from our source. HWP S changes the Schmidt
coefficients and HWPs A and B change the Schmidt basis; in
principle, this setup can produce any pure 2-qubit entangled
state with real coefficients. We vary $\theta_0$ and $\theta_1$
in steps of $15^\circ$ from 0 to $90^\circ$, resulting in 49
pairs of states for which we can demonstrate state
discrimination. We characterize each state by reconstructing
its density matrix from a tomographically over-complete set of
measurements $\ket{\SH}$, $\ket{\SV}$, $\ket{\SD} =
\frac{1}{\sqrt{2}}\left(\ket{\SH}+\ket{\SV}\right)$, $\ket{\SA}
= \frac{1}{\sqrt{2}}\left(\ket{\SH}-\ket{\SV}\right)$,
$\ket{\SR} =
\frac{1}{\sqrt{2}}\left(\ket{\SH}+i\ket{\SV}\right)$, and
$\ket{\SL} =
\frac{1}{\sqrt{2}}\left(\ket{\SH}-i\ket{\SV}\right)$ on each
photon with the the Pockels cell off. The states were
reconstructed using a semidefinite-program implementation
\cite{Doherty09} of the maximum-likelihood algorithm
\cite{James01}, and fidelities of the reconstructed density
matrices with their targets range from $0.97$ to $0.99$.

The optimal state discrimination measurement is the same for
all pairs of states of the form of Eq.~\ref{orthstates}. Alice
measures in the basis $\{\ket{\SH},\ket{\SV}\}$.  If she
registers outcome $\ket{\SV}$, Bob will measure in the
$\{|u\rangle, |u^\perp\rangle\}$ basis. On the other hand, if
she measures $\ket{\SH}$, the Pockels cell will fire on Bob's
side, changing his measurement basis to
$\{Z|u\rangle,Z|u^\perp\rangle\}$. We integrate coincidence
counts for 100s, monitoring all four combinations of outputs
($\mathrm{T}_a\mathrm{T}_b$, $\mathrm{T}_a\mathrm{R}_b$,
$\mathrm{R}_a\mathrm{T}_b$, $\mathrm{R}_a\mathrm{R}_b$). Here,
T(R) refers to detection in the transmitted (reflected) arm of
the PBS and the subscript $a$ ($b$) labels Alice's (Bob's)
analyzer.

Figs.~\ref{sampletomocounts}a)-d) show the real and imaginary
parts of the reconstructed density matrices for an example pair
of states $|\phi_0\left(30^\circ\right)\rangle$ and
$|\phi_1\left(60^\circ\right)\rangle$.  They have fidelities
$0.981\pm0.002$ and $0.986\pm0.002$ with the ideal states,
respectively. Ideally, if we prepare any
$|\phi_0\left(\theta_0\right)\rangle$ we should only detect
counts in the $\mathrm{T}_a\mathrm{T}_b$ and
$\mathrm{R}_a\mathrm{T}_b$ outputs, and if we prepare any
$|\phi_1\left(\theta_1\right)\rangle$ we should only detect
counts in $\mathrm{T}_a\mathrm{R}_b$ and
$\mathrm{R}_a\mathrm{R}_b$. In Figs.~\ref{sampletomocounts}e)
we show the experimentally measured counts for the states in
Figs.~\ref{sampletomocounts}a) \& b) (blue, left bars) and
Figs.~\ref{sampletomocounts}c) \& d) (red, right bars). From
these counts, we obtain the probability of correctly
determining the state; for example, the probability of
correctly determining $|\phi_0\left(\theta_0\right)\rangle$ is
given by
$P_0=\left(n_{\mathrm{T}_a\mathrm{T}_b}+n_{\mathrm{R}_a\mathrm{T}_b}\right)/\sum$,
where $n_{\mathrm{T}_a\mathrm{T}_b}$ is the number of counts
measured in the output $\mathrm{T}_a\mathrm{T}_b$, etc. and
$\sum$ is the sum of all four coincidence counts when
$|\phi_0(\theta_0)\rangle$ is prepared; a similar expression is
used for the probability of correctly determining
$|\phi_1\left(\theta_1\right)\rangle$, $P_1$. Using the counts
measured for the target states
$|\phi_0\left(30^\circ\right)\rangle$ and
$|\phi_1\left(60^\circ\right)\rangle$ shown in
Fig.~\ref{sampletomocounts}e), we find the average probability
for correctly determining the state is $P=0.9780\pm0.0003$. The
most significant difference from the theoretically perfect
discrimination can be explained by our imperfect initial state
fidelity.

For comparison, we consider a measurement strategy using local
projective measurements \emph{without} feed-forward, i.e.,
measurement in a basis
$\{\ket{\alpha},\ket{\alpha^\perp}\}\otimes\{\ket{\beta},\ket{\beta^\perp}\}$,
where $\ket{\alpha}$ and $\ket{\beta}$ are arbitrary
single-photon polarization states.  The optimal probability for
distinguishing between pairs of states was found by numerically
maximizing the success probability over all bases and all
possible ways of assigning measurement results. We assume each
state is produced with probability $0.5$. Without feed-forward,
the ideal states, $|\phi_0\left(30^\circ\right)\rangle$ and
$|\phi_1\left(60^\circ\right)\rangle$, can be distinguished
with only 0.933 probability and our experimental states, based
on our tomography, can be distinguished with $0.922\pm0.002$
probability. Our experimental measurement $P=0.9780\pm0.0003$
easily surpasses these limits, clearly demonstrating the
advantage feed-forward provides in state discrimination.

\begin{figure}
\begin{center}
\includegraphics[width=1 \columnwidth]{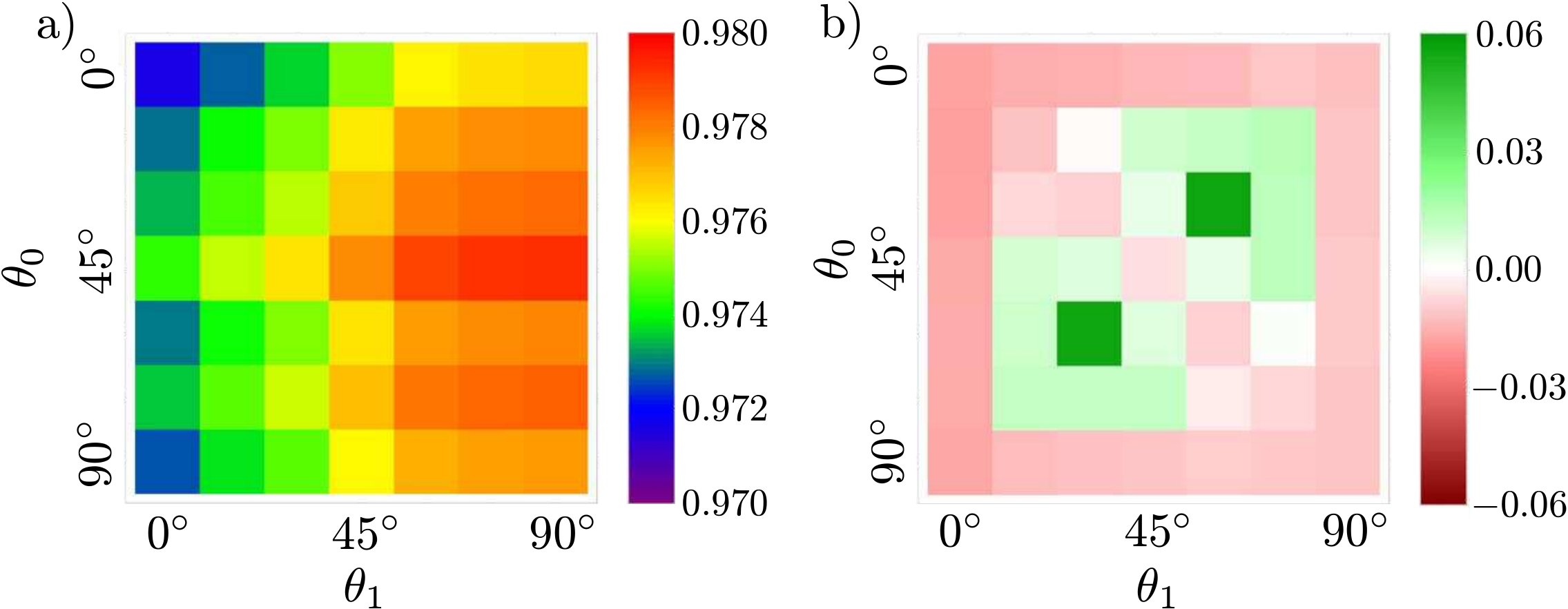}
\caption{Experimental minimum-error discrimination of
orthogonal entangled states.  a) Discrimination probability
for pairs of states
$|\phi_0(\theta_0)\rangle$ and $|\phi_1(\theta_1)\rangle$ for
$\theta_0$ and $\theta_1$ ranging from $0^\circ$ to $90^\circ$ in
steps of $15^\circ$ using local measurements and feed-forward.
All probabilities fall in the range from $0.9716$ to $0.9793$ with an average
uncertainty (1 sd.) of $0.0003$.  In principle, these
states can be discriminated perfectly by such a measurement, the
difference being mostly attributable to our state preparation fidelity.  b)
The advantage between the measured discrimination probability and the
optimal projective measurement \emph{without} feed-forward on our
experimentally produced states.  The variation in the values show that
some pairs of states are much easier to discriminate without
feed-forward than others. We measure an advantage up to $5.6\pm0.2$\%.
\label{summaryorth}}
\end{center}
\end{figure}

We summarize our results for the 49 pairs of states in
Fig.~\ref{summaryorth}.  In Fig.~\ref{summaryorth}a) we show
the measured discrimination probabilities for all pairs of
states, which all fall in the range from $0.9716$ to $0.9793$.
In Fig.~\ref{summaryorth}b) we show the difference between our
measured discrimination probability and the calculated optimal
local measurement based on our tomographically reconstructed
states. The maximum advantage our measurement achieved was
$5.6\pm0.2$\%. Note that in several cases we have a slight
disadvantage, up to $1.9\pm0.2$\%, as we are comparing our
experimentally measured probability with feed-forward to the
\emph{best possible} theoretical measurement without.

In the second part of our experiment, we use our setup to
perform minimum-error discrimination of equiprobable
\emph{non-orthogonal} entangled states of the form:
\begin{eqnarray}
|\psi_0\rangle & = & \cos ( 45^\circ\mbox{$-$}\eta ) |\phi_0(60^\circ)\rangle + \sin ( 45^\circ\mbox{$-$}\eta) |\phi_1(30^\circ) \rangle \\
|\psi_1\rangle & = & \sin ( 45^\circ\mbox{$-$}\eta) |\phi_0(60^\circ)\rangle + \cos ( 45^\circ\mbox{$-$}\eta) |\phi_1(30^\circ) \rangle \nonumber.
\label{nonorthstates}
\end{eqnarray}
Because these states have all real coefficients, we can prepare
them using HWPs S, A, and B. The overlap between the states is
$|\langle\psi_0|\psi_1\rangle|^2 = \cos^2 2\eta$. If
$\eta=\pi/4$, the states are orthogonal, and if $\eta=0$, the
states are identical. Our measurement can, in principle,
perfectly distinguish the states $|\phi_0(60^\circ)\rangle$ and
$|\phi_1(30^\circ)\rangle$ which corresponds to the
minimum-error, or Helstrom measurement for these pairs of
states when they are prepared with equal probability.

\begin{figure}
\begin{center}
\includegraphics[width=0.95 \columnwidth]{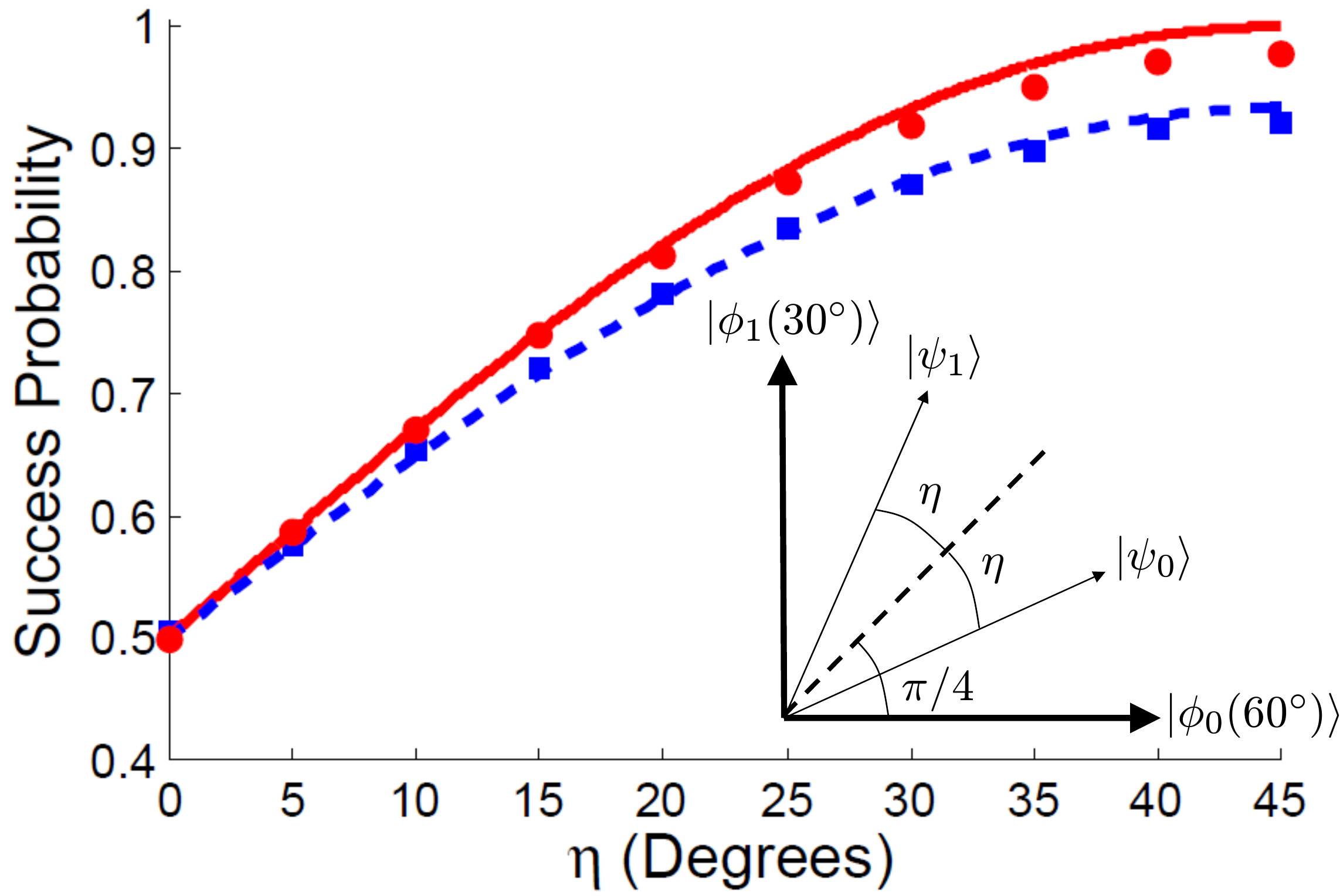}
\caption{Probabilities for successfully discriminating
pairs of non-orthogonal states. In the inset, we depict non-orthogonal
states $\vert\psi_0\rangle$ and $\vert\psi_1\rangle$ as superpositions
of $\vert \phi_0(60^\circ)\rangle$ and $\vert \phi_1(30^\circ)\rangle$ with the angle
$\eta$ characterizing their overlap. The experimentally measured
probabilities for distinguishing $\vert\psi_0\rangle$ and
$\vert\psi_1\rangle$ (red circles), with errors, $0.0007$, too small too see, and theoretical prediction
given ideal state preparation (red solid line) are shown as a function
of $\eta$. The data
follow the prediction closely, taking the initial fidelity into account. The best possible discrimination probability
by local projective measurements without feed-forward
is shown for the ideal states (blue dashed line) and from the tomographically
reconstructed states (blue squares).  Our results
show a distinct advantage arising from feed-forward. \label{fig_nonorth}}
\end{center}
\end{figure}

Fig.~\ref{fig_nonorth} shows the experimentally measured
probability for distinguishing pairs of non-orthogonal states
(red circles) as a function of the overlap parameter $\eta$. We
observe good agreement between the data and theoretical
expectation (red solid line). The ratio of the experimental
results to the theoretical expectations ranges from
$99.9\pm0.2$\% for nearly identical states ($\eta$=$0$) to
$97.65\pm0.03$\% for nearly orthogonal states
($\eta$=$45^\circ$); again, this difference is mostly due to
experimental state preparation fidelity. The probabilities for
the best possible discrimination using local projective
measurements without feed-forward are shown in
Fig.~\ref{fig_nonorth} for the experimentally produced states
(blue squares) reconstructed via tomography and for the ideal
states (blue dashed line). Our experimental state
discrimination results show a clear advantage over both of
these quantities.

Optimal quantum state discrimination plays an important role in
quantum information.  We have implemented optimal minimum-error
discrimination of orthogonal and non-orthogonal entangled
quantum states using only local measurements and feed-forward.
Our experimental results show a clear advantage over the best
possible projective measurement without feedforward.  This
demonstrates a rather surprising ability of local measurements
to distinguish global properties, i.e., entanglement, and is
the first implementation of a state discrimination protocol in
a uniquely quantum regime.

The authors gratefully acknowledge financial support from
NSERC, QuantumWorks, MRI ERA, the Perimeter Institute and CFI.
Research at the Perimeter Institute is supported by Industry
Canada and by the Ontario MRI.

\end{document}